# Independent publishers and social networks in the 21st century: the balance of power in the transatlantic Spanish-language book market

## Authors

**Ana Gallego-Cuiñas** E-mail: [anag@ugr.es](anag@ugr.es)

**Esteban Romero-Frías** E-mail: [erf@ugr.es](erf@ugr.es)

**Wenceslao Arroyo-Machado** E-mail: [wences@ugr.es](wences@ugr.es)

## Abstract

**Purpose -** The present paper uses Twitter to analyze the current state of the worldwide, Spanish-language, independent publishing market. The main purposes are to determine whether certain Latin American Spanish-language independent publishers function as gatekeepers of World Literature and to analyze the geopolitical structure of this global market, addressing both the Europe-America dialectic and neocolonial practices.

**Design/methodology/approach -** After selecting the sample of publishers, we conducted a search for their Twitter profiles and located 131; we then downloaded data from the corresponding Twitter APIs. Finally, we applied social network analysis to study the presence of and interaction between our sample of independent publishers on this social media.

**Findings -** Our results provide data-based evidence supporting the hypothesis of some literary critics who suggest that in Latin America, certain publishers act as gatekeepers to the mainstream book market. Therefore, Twitter could be considered a valid source of information to address the independent book market in Spanish. By extension, this approach could be applied to other cultural industries in which small and medium-sized agents develop a digital presence in social media.

**Originality/value -** This paper combines social network analysis and literary criticism to provide new evidence about the Spanish-language book market. It helps validate the aforementioned hypothesis, proposed by literary critics, and opens up new paths along which to pursue an interpretative, comparative analysis.

# 1. Introduction

Publishing, one of the most important culture-based industries, has been markedly affected by the remarkable rise of independent publishers in the 21st century—a phenomenon that has expanded both in Europe and North and Latin America. Independent publishers have to a certain extent counteracted the strong monopolistic tendencies of a sector in which—since the expansion of the global market—the major firms have bought out any smaller, supposed competitors that might threaten their economic growth.

In this context, independent Spanish-language publishing houses, the object of the present study, are well worth in-depth study. Authorial and discourse practices are constructed as a consequence of the modes of production and of distribution of these small- and medium-sized publishers, whose publications sometimes gain access to the major publishing markets. Likewise, the way companies interact within the ecosystem of independent publishing facilitates a decolonial reading (Mignolo, 2009), given the cultural diversity that favors not only a more plural market, but new 21st-century hegemonies between Latin American countries and Spain.

The present study aims to analyze independent Spanish-language publishers by applying social network analysis to the presence of and interaction between their profiles on Twitter. This Digital Humanities (DH) approach provides us with evidence of how they interact and this is structured in quantifiable data that enables us to compare the independent publishers with the findings of traditional research in the sociology of culture (Williams, 1983; Bourdieu, 2002).

## 1.1. Independent publishers in the 21st century

Independent publishers are one of the outstanding features on the 21st-century landscape of cultural entrepreneurship in North and South America and Europe. Furthermore, they constitute a highly-favoured object of study in the human and social sciences (Schiffrin, 2011; Szpilbarg and Saferstein, 2012; López Winne and Malumián 2016; Gallego Cuiñas and Martínez, 2017; Hawthorne, 2018; Guerrero, 2018; Locane, 2019). Within the book industry, the proliferation of this business model has opened up new, alternative, sustainable literary markets that counteract the obsolescent logic of the large monopolies (Sapiro, 2009; Thompson, 2012; De Diego, 2015). The small and medium-sized companies' search for commercial opportunities that larger firms have ignored is the key to their success with the culturally-oriented public as they showcase new writers, minor genres (Deleuze and Guattari, 1978) such as poetry, theatre and essay, feminist and LGTBQ works, uncatalogued works, and so on. In just a few years, a good number of these companies have built up rich, diverse, daring, avant-garde catalogues that rival the hegemony of symbolic value of their economically more powerful competitors.

Despite the apocalyptic forecasts of the death of the printed book, to be replaced by the electronic format (ebook) (Davis, 2015), more and more is being published on paper and the number of new publishing houses is constantly growing. This is partly due to the boom in independent publishers in the 21st century. Several factors have influenced this situation: firstly, the technological developments (e.g. simple layout programs, cheaper design and printing) and improvements in the conditions in which books are physically produced; secondly, the adoption of innovative business models that take advantage of digital technology's potential (Celaya, 2014; Celaya et al., 2015); thirdly, the economic crisis that reduced businesses to self-employed one-man-bands, surviving on print editions and self- or local distribution to a highly specialized readership; and finally, the market domination of major publishing groups, which began in the 1980s (Sapiro, 2009), and led to the progressive marginalization of less profitable genres and authors, so that unattended niche markets could be occupied by small independent companies.

The Spanish-language publishing industry is controlled by two transnational conglomerates: the Berltersmann group, backed by German capital, and the Planeta group, with Spanish capital. Consequently, the modes of production and distribution of the symbolic capital pertaining to the Latin American cultural space have been deterritorialized as they are articulated through a Eurocentric "coloniality of power" (Mignolo, 2009) that imposes its hegemonic evaluation of literary production which, for example, favors the novel and traditional aesthetics. Truth to tell, for Latin American literature this Eurocentric neocolonial exploitation of its cultural riches is a heavy burden to bear. The two publishing conglomerates mentioned earlier are highly competitive on a transnational scale although they also create national ghettos which make it impossible to distribute anything less than a "best seller" in Central and Southern America. This dynamic reproduces core-periphery inequalities (Wallerstein, 1991) dating from the colonial era, since Spain remains the economic and symbolic epicenter of Spanish-language book distribution, to the point that Latin American writers are highly unlikely to be read in Europe unless their work has been published in Spain.

However, since the early 21st century, we have witnessed an authentic boom in small- and medium-sized independent Spanish-language publishers in Latin America (Gallego Cuiñas, 2017, 2018), which represent an emancipated (Rancière, 2014), decolonial (Mignolo, 2009) channel of production. So, what may have started out as a local strategy to accumulate economic capital in niche markets abandoned by larger publishing groups, has become a means of producing symbolic capital pertaining to South American (De Sousa Santos, 2010), supporting genres like the short story and essay, and new writers employing more avant-garde aesthetics. Paradoxically, though, this cultural niche production must later be absorbed by the larger groups in order for it to reach an international market. Thus, independent publishers play a fundamental role in protecting bibliodiversity and guaranteeing heterogeneity and plurality in the ecosystem of the book. Moreover, they are responsible for the decolonial cultural value of World Literature (Damrosch, 2003) in 21st century Spanish. Independent publishers use current-day digital communication tools to survive, which is why they generally develop an active presence on social networks. In the Art world, this process has been analyzed in depth (Heinich, 2004; González Martín, 2018), but in the field of literature, research has used Bourdieu's Field theory to focus on the sociology of the networks woven by mediators and agents. Notwithstanding, we cannot ignore the fact that sociability networks in the field of literature are built on new technologies such as Twitter. As Boll (2011) points out, on Twitter network identity is presented as a fundamental management tool that generates trust and value, as well as developing exchanges and a lasting digital reputation.

## 1.2. Social network analysis in the Digital Humanities

The term Digital Humanities (DH) first appeared in *A companion to digital humanities*, edited by Schreibman, Siemens and Unsworth (2004). It represented a contemporary name for research in the Humanities using digital technology to answer classical questions and address new issues raised by the digital world. A key element of DH is the incorporation of large-scale data analysis to answer questions that were previously addressed via micro and qualitative approaches (Jockers, 2013). Examples abound, ranging from the use of digitalized text analysis to provide in-depth literary studies to the social network analysis of authors, their works, and so on (Moretti, 2006; De Nooy et al., 2005; Michel et al., 2011; Wilkens, 2015; Martínez-Gamboa, 2015).

When the object of study corresponds to current problems reflected in social network use, we can analyze the information published and the relationships established between agents in online spaces (Zaheer et al., 2010). For this type of analysis, Twitter is one of the most widely-used networks since it has important advantages: the openness of its communications; the opportunity to maintain non-reciprocal relationships in which one profile can follow another without establishing a mutual link; and the provision of an API to extract data in a structured, automated way, among others.

Hence, Twitter has been used to analyze relationship or discussion networks between agents in subjects related to the Humanities and Social Sciences: e.g. the global community of DH researchers (Grandjean, 2016), political institutions like the European policy labs (Romero-Frías and Arroyo-Machado, 2018), gender and issue discussion between women (Evans, 2016) or online campaigning and political involvement (Kruikemeier et al., 2016). Elsewhere, research has focused on dialogue arising from specific events such as international academic conferences (Grandjean and Rochat, 2014; Jussila et al., 2014) social events (Beguerisse-Diaz et al., 2014; Casilli and Tubaro, 2012) or political issues (Lewis et al., 2019). The profiles of agents belonging to Twitter communities have also been analyzed: e.g. in the case of DH centers (Romero-Frías and Del-Barrio-García, 2014). Consequently, we would affirm that the study of these relationships also allows us to establish indicators of the influence of agents in a given community (Subbian and Melville, 2011), for example in the case of science studies (Haustein et al., 2014) or political communication (Stieglitz and Dang-Xuan, 2012).

# 2. Objective and research questions

Our objective is to use Twitter to analyze the current state of the worldwide, Spanish-language independent publishing market. To do so, we start with two research objectives: firstly, we aim to test whether certain independent Spanish language Latin American publishers function as gatekeepers of World Literature. The term 'gatekeeper', applied to Literary Studies, was popularized by Loren Glass in the 1970s and refers to literary experts who regulate the relationship between author and audience, promoting certain works and creating writing and reading habits (2008). More recently, Marling (2016) carries out a detailed analysis of the "interaction rituals" of writers (e.g., group formation, promotions, challenging strategies, etc.), and of the gatekeepers of world literature since the 1960s: specific characters—publishers, translators, writers—who have acted as mediators for other authors. However, our position in the present study is closer to that of Thompson (2012), who considers that gatekeepers function as a device (Foucault, 2012) that gives visibility to and promotes Latin American authors who are later published and translated by the large international groups (Gallego Cuiñas, 2018).

Secondly, we would ask about the geopolitical structure of the global independent Spanish-language publishing market, addressing the Europe-America dialectic and neocolonial practices in this market. The geopolitical factor defines the modes of editorial production in Latin America—which is economically dominated by European capital—which is why we make a joint comparison between the main countries of the subcontinent and Spain.

Our objective then will be to answer these two questions by applying a quantitative method based on social network structure analysis (Myers et al., 2014; Grandjean, 2016) of a large corpus of independent publishers in North and South America and Spain, based on the Twitter follower relationships established between them. The study of social networks has become increasingly important in social studies as a favored means of understanding the nature of the professional relationships and cooperation between entrepreneurs, artists and institutions (De Nooy, 1991). Independent publishers are particularly active on social media in order to gain visibility. By following a publisher profile, a user can establish a clear picture of how social and material relations are established in the independent book market, the works they publish, and the writers they promote. As we pointed out above, the independent nature of publishers is also a determining factor in the significant use of networks, since by not having large resources for the commercial promotion of their books, they must use social networks skillfully in order to promote their publications and contact potential readers and writers.

# 3. Methodology

## 3.1. Selection and sample of independent publishers

To carry out our research into the behavior of independent Spanish-language publishers on Twitter, we have used data obtained from the "Independent publishers for the literary ecosystem" (www.ecoedit.org)—the ECOEDIT project—funded by the University of Granada, Spain.

The choice of independent Spanish-language publishers was based on two primary sources: the *Alliance internationale des éditeurs indépendants* and EDI-RED, the Portal of Ibero-American Editors and Publishers, hosted in the Miguel de Cervantes Virtual Library. We also consulted country-specific secondary sources such as publishers' associations, independent publishers' networks and groups, attendees at book fairs and publishers' fairs, and publishers' catalogues. Note that to date, in no single space whether public or private, national or transnational, do all independent Spanish-language publishers converge; hence we have consulted a range of sources. To do so, we constructed a representative, non-probabilistic sample of 200 independent publishers from different countries. In 2018, a survey was sent to those included and we received 163 answers. The publishers in our sample have much in common: they are not exclusively dedicated to self-publishing for profit; they are currently active with at least one book published in 2018; and they are private companies. Moreover, over 90% publish with an ISBN—the exceptions are the so-called "Cartonera Collectives" and some linked to cultural associations.

We searched for the Twitter profiles of the 163 respondents via Google, the Twitter search engine, and links on their institutional web pages. In December 2018, we found 131 active profiles (80% of the total), which confirms that their Twitter activity is representative of the sector. Our final sample included the major independent firms, allowing us to analyze how the sector is structured through their online connectivity.

## 3.2. Twitter data collection and processing

Twitter data was downloaded for each of the accounts found by accessing the Twitter API, using a script programmed in Python and the Tweepy library. At December 31, 2018, we obtained descriptive values of the profiles and complete lists of followers (users who follow an account) and followings (users followed by an account) for each account.

The publishers' network was analyzed by visualizing the connections between the 131 Twitter profiles by using Gephi (Bastian et al., 2009). Analysis was conducted both globally, paying attention to the entire network, and locally, studying the role of each node. At a global level, we calculated the diameter and the average distance between all of them. For each node, we took account of the degrees of input (indegree) and output (outdegree). These concepts are equivalent to Twitter account followers and followings, respectively. Similarly, we calculated the centrality measures of degree, betweenness and eigenvector. The first shows the total number of input connections that a node has received; the second indicates the role a given node plays as a bridge within the network—i.e. if it frequently appears on short paths (the minimum path that can be established to connect two nodes) among all node pairs. Eigenvector centrality is a better measure than degree to understand the relevance of nodes in a network since it not only quantifies the number of followers but also their relevance, an approach similar to the way that Google's Pagerank provides relevant search results. We also calculated the clustering coefficient to learn the extent to which they were connected to each other.

# 4. Results

## 4.1. Presence of independent publishers on Twitter

Of the 15 countries in the sample, those with greater editorial presence are: Spain (29 of 163, 72.5% of all publishers based in Spain), Argentina (23, 57.5%), Chile (19; 95%), Mexico (16; 53.3%), and Colombia (14; 70%).

Quantitative data from the Twitter profiles can be consulted in Supplementary Table 1. The 131 publishing houses have mean figures of 7838.21 followers (SD: 23 034.22), 1379.6 followings (SD: 2560.57), and 5233.53 tweets (SD: 10 719.63).

By followers—an indicator of the relevance of presence in the network (Zarco et al., 2016)—the top 5 publishers are: *Nordica_Libros* (237 076), *EdSextoPiso* (57 659), *LibrosAsteroide* (55 080), *eternacadencia* (48 472), and *EdImpedimenta* (45 609).

The profile creation date indicates when each publisher began using Twitter to promote and communicate its activities. The two earliest profiles belong to: *BlattyRios* (January 7, 2008) and *lapollera* (April 6, 2008). Over the years, profile creation is distributed as follows: 12 new profiles in 2009, 28 in 2010, 26 in 2011, 17 in 2012, 17 in 2013, 8 in 2014, 10 in 2015, 5 in 2016, 2 in 2017 and 4 in 2018.

By follower/following ratio—interpreted as an indicator of influence on Twitter—the top 5 publishers are: *Nordica_Libros* (301.6), *irrupcionesge* (124), *EdSextoPiso* (66.9), *eternacadencia* (66.1), and *MANSALVA_* (57.8). Twenty-five publishers have ratios of less than 1, indicating that they either use Twitter as a means of tracking information or they have minimal relevance in the community.

The mean number of followers of these 25 accounts is 649.1, lower than the global average. The corresponding number of published tweets is similarly low: 762.3. Some 77 publishers have a ratio over 2, which indicates that they have at least twice as many followers as followings. In line with Grandjean (2016), we are dealing with publishers who use Twitter as a kind of technological monitor, in the first case, and with "stars", in the second.

The number of tweets posted, including retweets, serves as an indicator of publisher activity on Twitter. The most prominent publishers are: eternacadencia (89 463), OlgaCartonera (53 923), LectorComplic (36 405), MalpasoEd (30 546) and Catalonialibros (24 832).

On the other hand, we must also consider the degree to which accounts are current. The vast majority (116 accounts, 88.55% of the total) have published at least one tweet in 2018. The least current accounts are: *ayarmanot* (last updated in 2012), *TallerEdicion* (2012), *lapixeditores* (2013), *puenteaereo_ed* (2013) and *ProyectoVOX* (2014).

## 4.2. Cartography of the independent publishing sector in Spanish

The sociability network of Twitter, built from the following relationships that publishers (the network nodes) establish among themselves, consists of a core of 125 of the 131 publishers. Furthermore, this high degree of interconnection is clearly reflected in the average clustering coefficient (0.433) and the low average distance (2.26) and diameter (6). The 6 publishers excluded from the core are: *AletheyaPeru*, *ayarmanot*, *editorialDR*, *EstruendomudoPe*, *FindeSiglo1* and *lapixeditores*.

At node level, the basic indicator is degree, which can represent input connections (indegree, followers on Twitter) and output connections (outdegree, followings on Twitter). Other important measures are degree centrality, betweenness centrality and eigenvector centrality. Degree centrality shows a given node's total number of input connections; betweenness centrality indicates the role that a node plays as a bridge within the network—that is, whether or not it appears frequently on short paths (the minimum path connecting two nodes) with respect to all node pairs. Eigenvector centrality enhances this since it quantifies both the number of followers and their relevance providing relevant search results in an approach similar to that of Google's Pagerank.

Figures 1 and 2 show the internal connections network between the publishers in the sample. Node size indicates the indegree (number of followers) on the basis of the nodes within the network of publishers and excluding any followers outside of this but belonging to the Twitter network as a whole. Figure 1 represents each publishers' country of origin by colour coding: Spain (blue, 23.2% of the sample), Argentina (orange, 17.6%), Chile (pink, 15.2%), Mexico (green, 12%), Colombia (brown, 11.2%), and all other countries (grey, 20.8%).

Figure 1. Map of internal connections (node size reflecting indegree, colour coding by country).

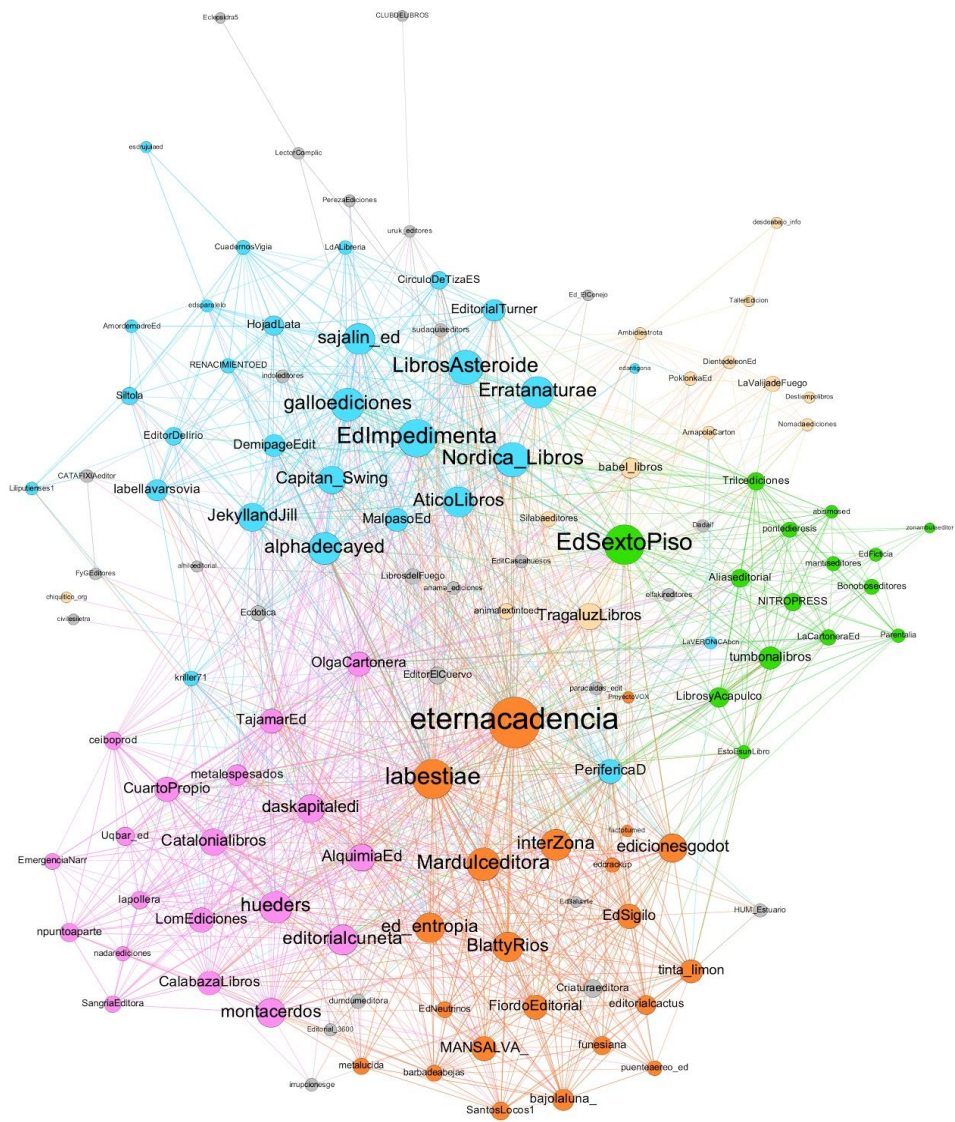

Since our sample only includes Spanish-language publishers, we cannot identify linguistic patterns; however the geographic component is of great relevance. Figure 1 shows how the 4 countries with the highest percentage of publishers in the sample make up clusters. Chilean and Argentinian publishers constitute the group with the greatest geographical and cultural proximity. Spain and Mexico make up two well-defined spaces closer to each other than they are to the Southern Cone publishers.

Table 1 ranks the main publishers in terms of centrality measures that include indegree, eigenvector centrality and betweenness centrality in the network. Centrality enables us to quantify the importance of the nodes in the network and identify the key actors (Newman, 2010). Indegree indicates the number of followers for each publisher within the network. Eigenvector centrality indicates the relevance of any given publisher within the network.

Finally, betweenness shows how often a node is present on the shortest path between any pair of vertices in the network, allowing us to identify which nodes function as bridges between network clusters.

These instruments allow us to address our first research question: Do certain independent Spanish-language Latin American publishers function as gatekeepers of World Literature? In this context, Eterna Cadencia is in a remarkable position as it is ranked first on all three indicators: indegree (77), eigenvector centrality (1) and betweenness (2193.59). This represents the recognition its publications receive from the rest of the network. This publishing house plays a significant role as a mediator for avant-garde writers and aesthetics that are later adopted by the large groups, as evidenced by writers like Gabriela Cabezón Cámara or Vera Giaconi who, after having been published by Eterna Cadencia, appeared in Random House and Anagrama Literature, respectively.

Eigenvector centrality shows us that although 5 of the top 10 publishers are based in Spain, 3 of the top 5 are based in America: Eterna Cadencia, as we have already mentioned, labestiae (Argentina, 2nd), and EdSextoPiso (Mexico, 4th). Betweenness centrality reveals that only one Spanish publishing house ranks in the top 10, Periférica (7th).This particular publisher clearly tries hard to link up with Latin America since it distributes in Argentina, Mexico, Peru and Colombia and publishes many contemporary Spanish American authors (Yuri Herrera, Carlos Labbé, Juan Cárdenas, Diamela Eltit, Indiana Rita, Maximiliano Barrientos and Nicolás Cabral, among others). Betweenness centrality also explains why Periférica follows some 59 independent Latin American publishers but only 13 Spanish national publishers, including those within their own Contexto group. Finally, apart from Eterna Cadencia, the highest ranking positions for betweenness centrality are taken by Spanish American publishers, with *OlgaCartonera* (Chile, 2nd) and *labestiae* (Argentina, 3rd) being the most noteworthy.

We also found that some Latin American publishers act as links between national publishing communities: for example *OlgaCartonera* or *EdSextoPiso*, the latter having its headquarters both in Mexico City and Madrid. Independent Spanish publishers play a less evident connecting role even though the eigenvector centrality indicator shows they do hold positions of relevance in the network. This may be indicative of the fact that Latin American publishers view their Spanish counterparts as a market for expansion and a bridge to the large European publishing groups.

Table 1. Rank table of the main independent publishers by indegree, eigenvector centrality and betweenness centrality

| | indegree | | | eigenvector centrality | | | betweenness centrality | |
|---|---|---|---|---|---|---|---|---|
| 1 | eternacadencia (Argentine) | 77 | 1 | eternacadencia (Argentine) | 1.0 | 1 | eternacadencia (Argentine) | 2193.59 |
| 2 | EdSextoPiso (Mexico) | 56 | 2 | labestiae (Argentine) | 0.79 | 2 | OlgaCartonera (Chile) | 1895.41 |
| 3 | labestiae (Argentine) | 55 | 3 | EdImpedimenta (Spain) | 0.73 | 3 | labestiae (Argentine) | 1428.51 |
| 4 | EdImpedimenta (Spain) | 52 | 4 | EdSextoPiso (Mexico) | 0.65 | 4 | TragaluzLibros (Colombia) | 935.77 |
| 5 | LibrosAsteroide (Spain) | 46 | 5 | Nordica_Libros (Spain) | 0.65 | 5 | EdSextoPiso (Mexico) | 530.47 |
| 6 | Nordica_Libros (Spain) | 46 | 6 | LibrosAsteroide (Spain) | 0.65 | 6 | LibrosdelFuego (Venezuela) | 528.67 |
| 7 | galloediciones (Spain) | 43 | 7 | alphadecayed (Spain) | 0.63 | 7 | Periférica (Spain) | 528.07 |
| 8 | Mardulceditora (Argentine) | 43 | 8 | Mardulceditora (Argentine) | 0.62 | 8 | daskapitaledi (Chile) | 526.22 |
| 9 | alphadecayed (Spain) | 41 | 9 | galloediciones (Spain) | 0.59 | 9 | JekyllandJill (Chile) | 435.79 |
| 10 | Erratanaturae (Spain) | 41 | 10 | hueders (Chile) | 0.59 | 10 | hueders (Chile) | 360.47 |
| 11 | hueders (Chile) | 41 | 11 | Erratanaturae (Spain) | 0.59 | 11 | EdImpedimentac (Spain) | 324.01 |

## 4.2. Analysis of the transatlantic network of relations between independent publishers in Spain and Latin America

To analyze the relationships between the publishers in Spain and Latin America, we examined all the connections between the 29 Spanish and 102 Latin American publishers. Table 2 shows that of the 464 following relationships Spanish publishers establish with other publishers in the sample, 315 (67.89%) are with Spanish publishers, while 149 (32.11%) are with Latin American publishers. Of the 1592 connections that are established from Latin America, 80.15% are with other American publishers, while only 19.85% are with Spanish publishers.

Table 2 . Connections between independent publishers' Twitter accounts between Spain and Latin America

| Following\Follower | Spain | Latin America |
|---|---|---|
| Spain (464 connections) | 315 (67.89%) | 149 (32.11%) |
| Latin America (1592 connections) | 316 (19.85%) | 1276 (80.15%) |

Figure 2 breaks down the relationships between the main countries by number of publishers. The horizontal axis shows from which countries publishers establish a connection to their counterparts in the other countries named along the vertical axis. For instance, there are 315 connections between publishing houses based in Spain and 117 connections between Chilean and Argentinian publishers.

Figure 2. Connections between publishers in different countries.

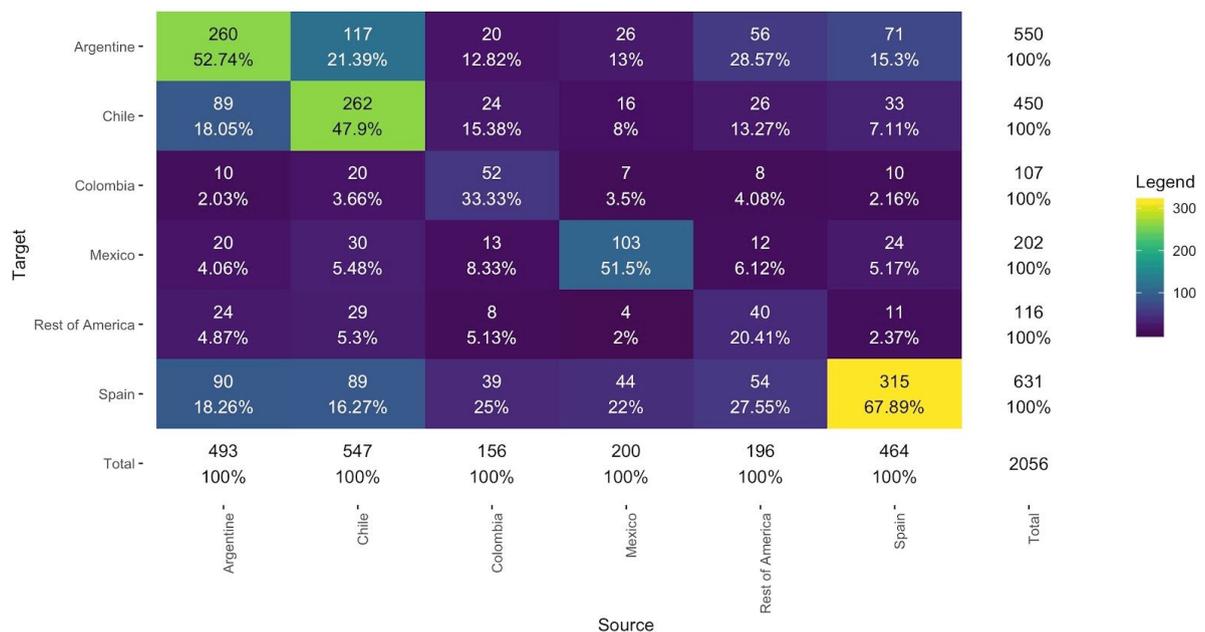

If we look at the number of connections established by each country, in absolute terms, Chile with 547 (47.9% of which are with other Chilean publishers) ranks first, followed by Argentina with 493 (52.74%), Spain with 464 (67.89%), and Mexico with 200 (51.5%). Relatively speaking, if we take account of the number of publishers in each country, we found that those who are most active in establishing connections are: Chile with 28.8 connections per publisher; Argentina, 21.4; Spain, 16; and, Mexico, 12.5.

Among the Latin American countries, Mexican publishing houses are those that most clearly look towards Spanish publishers, 45.36% of its links, followed by Argentina (38.64%), Colombia (37.5%) and Chile (31.23%). The explanation for this lies in the fact that Mexico has a less robust, more controlled publishing industry than its Southern Cone peers (Argentina and Chile), where independent publishers have proliferated since 2010 due to the implementation of both public and private policies leading to a more self-sufficient, interconnected model. This has created a geographical cluster, from which Mexico is excluded, thus being distanced from its Latin American counterparts. This is also supported by Mexico's desire to continue building transnational bridges with Spain which it sees as the headquarters and gateway to publishing and translation by large publishing groups.

## 5. Discussion

Of the 163 publishers included in the initial sample, 80% had a Twitter profile, which shows publishers' high level of use of this online communication network to promote their activity. The dates of creation of these profiles show how, in many cases, their digital sociability has been developed over quite some time. An indicator of the type of use is also the degree to which they update their accounts: the vast majority (116 accounts, 88.55% of the total) have published at least once in 2018. This, together with the results of our social network analysis, indicates that Twitter is a viable path along which to conduct future research on the publishing sector. The independent publishing community constitutes a small world (Watts and Strogatz, 1998) composed principally of 125 interconnected publishers. So this mutual recognition of agents within the independent publishing industry serves as their means to identify themselves.

In response to our question about which independent publishers act as gatekeepers to World Literature in Spanish, Figure 1 illustrates the fact that although Spain is the country with the highest percentage of publishing houses over the total, the most central publishers are Eterna Cadencia (Argentina), followed by Sexto Piso (Spain-Mexico), and La Bestia Equilatera (Argentina). Our visualization of this social network shows both Eterna Cadencia's high number of followers and its relevance in terms of its links with other prestigious actors.Eterna Cadencia clearly plays a cardinal role in the transatlantic map of independent publishing as a mediator and gatekeeper of World Literature in Spanish. Furthermore, the three central publishers (Eterna Cadencia, Sexto Piso and La Bestia Equilatera) have distributors—head offices even—in Spain, which is the cause and consequence of both their prominence on the social networks and their role as transatlantic bridges.In this, they contrast with their Spanish peers who—with the exception of Periférica—are less visible in Latin America. In fact, the aforementioned three publishing houses share a hybrid business model that moves between small, local, craft publishers with highly avant-garde, resilient production lines, and the use of stable, internationally competitive marketing channels like those of the major groups. The role played by Eterna Cadencia, ranked first on all three indicators, confirms the hypothesis posited by Gallego Cuiñas (2018). In her earlier study she pointed out that publishing houses such as Eterna Cadencia in Buenos Aires (Argentina) select, produce and distribute literary works oriented towards Argentina in certain genres (story and essay), with alternative, resistant, counter-hegemonic aesthetics. These are later taken up by major publishing groups such as Random House or Planeta and, even, Spain's Anagrama. These independent publishers, which act as gatekeepers to the Latin American poetic world, also create audiences with different tastes (Spivak, 1994) that are legitimated in other ways through literary criticism, cultural journalism, prizes, cinema, and other independent publishers.

Likewise, Eterna Cadencia's indegree and eigenvector centrality indicate its large social capital. We must not ignore the importance of the social networks to producers and consumers of culture, whose nature is transformed through technological mediation and the Internet, which has benefited peripheral actors, such as those in Latin America. The two publishing houses with highest eigenvector centrality in our sample are the Argentinian, Eterna Cadencia and La Bestia Equilatera, followed by the Spanish Impedimenta and the Mexican-Spanish Sexto Piso. They hold the leading positions in terms of social capital. Spain dominates the top 10 (except for the Argentinian Mardulce, and the Chilean Hueders), which confirms the importance of economic development—clearly much greater in Europe—for the book industry. The case of Argentina's Eterna Cadencia is not common in Latin America (it is only comparable to Sexto Piso, the second Latin American gatekeeper). Its substantial financial capital shows that other power relationships can be described in transatlantic publishing. These are based on combining local modes of production—in contrast to the deterritorialization that comes with the decolonial practices of the large groups—that converge to construct a unique editorial perspective. This catalogue is oriented towards language use as well as towards plots and themes (again, unlike those of the large groups) and heavily focused on national Latin American authors (more than half). The diversity of genres is also highly significant. For instance, the essay represents 45% of the catalogue, showing a strong commitment to critical thinking and to training active–subversive readers instead of passive–evasive consumers.

Generally, the results of eigenvector centrality (with a relatively balanced presence of Latin American and Spanish publishers in the top 10) and betweenness centrality (with a majority presence of Latin American publishers) suggest that a panorama exists in which relevance in the network of the main independent publishers is shared between Spain and Latin America. Nevertheless, Latin American publishers are playing a remarkably more active role in connecting both markets and building connections with the European environment, where the large worldwide publishing groups are settled. As we have pointed out, in Spain only Periférica evidences a significant betweenness role, derived from its distribution and publication of some of the most relevant contemporary Latin American authors. Thus, dominance of the independent Latin American publishing market and the possible construction of its own distribution space, combines with a transatlantic view that serves as a bridge to Spain as a gateway to the great European publishing market.

In relation to our second research question—relating to the geopolitical relationships between Latin American and Spanish publishers, we confirmed the existence of a large, "disobedient" Latin American symbolic capital (Mignolo, 2009) that is currently being produced by independent publishers. This could be considered a decolonial cultural practice since these companies follow an emancipated business model, with its own financial capital and produce catalogues that opt for less commercial genres (stories and essays) and more avant-garde aesthetics than those reproduced by large North American transnational groups. This subverts the economic and symbolic hegemony that Spain has held over the market for Spanish-language cultural products. Other hegemonies, clearly represented by Eterna Cadencia and Sexto Piso, are created in Spanish-speaking South America and followed by large groups and some Spanish independent publishers, such as Periférica. So, for the first time in the last 50 years of publishing house relationships between Latin America and Spain (De Diego, 2015), modes of production exist in the Latin American market that are not deterritorialized (to Spain or Germany). Consequently, Latin America is now culturally emancipated from Spain, although it has in no way turned its back on its transatlantic counterpart. Furthermore, these conditions (e.g. the close relations between publishers in the Southern Cone) favor the creation of a Latin American publishing network, with its own market, in which the industry itself controls the modes of production and of distribution.

Finally, we must take into account the connections between Latin American and Spanish publishers, illustrated through the data in Table 2. Although these two heterogeneous political realities are not comparable due to their size and diversity, their use of these connections is crucial because a decolonial reading of publishing practices forces us to confront the former metropolis, which still possesses a large economic and cultural capital. The fact that the Latin American publishers as a whole more closely follow their geographic neighbors (80% of the connections) than their Spanish peers (20%) can be interpreted as meaning that Latin American publishing houses are less interested in Spanish publishers which, with the exception of the conglomerates, are generally losing their hegemony.

However, to mitigate the effect of size in this comparison (Latin America vs. Spain), Figure 2 shows data by country. Spain is the most inward-looking country (67.89% of connections), compared to Argentina, Chile and Mexico where the percentage of internal connections approaches 50%. If we look at the number of connections established by each country, in absolute terms, Chile comes top with 547. The Chilean book market has developed considerably in recent decades, only comparable to that of Argentina, with prominent independent publishing houses such as Hueders and Laurel. Of all the Latin American countries, Mexico looks towards Spanish publishers the most with 45.36% of its links, followed by Argentina (38.64%), Colombia (37.5%) and Chile (31.23%). As we have already pointed out, this is motivated by the geographical distance of Mexico from the other Latin American countries—which hinders contact and distribution—and by Mexico's historically close relationship with the Spanish book market (De Diego, 2015), as illustrated by Sexto Piso. Given this, Mexico is located in an intermediate position between South America and Europe, with an independent publishing industry that is closer to that of Spain, as was also shown in Figure 1. These data reinforce the idea that, on the Latin American map of independent publishers, Argentina and Chile take a more emancipated, cultural and economic position, with respect to Spain. Furthermore, their physical closeness propitiates the creation of a singular, strong common space for editorial exchange in the Southern Cone, which has grown through the start-up of co-editions (such as those carried out by Eterna Cadencia), their own distributor (Big Sur) and the FILBA Literary Festival (https://filba.org.ar/) which alternates between Montevideo, Buenos Aires and Santiago, initiatives developed by Pablo Braun, the owner of Eterna Cadencia.

# 6. Conclusion

Social capital has always been an influential factor in the publishing sector. Eterna Cadencia holds the central position in our social network analysis of Twitter as indicated by eigenvector centrality and betweenness centrality data. This leads us to conclude that this Argentinian publisher has high social and symbolic recognition among the other small- and medium-sized Spanish-language publishers.

In the present study, the visualization of the publishers' network has been crucial in answering our research questions relating to the new hegemonies in the Spanish-language publishing sector. Thus, the hypothesis that Eterna Cadencia acts as a gatekeeper to Latin American literature is confirmed by its central position in this network. This reveals its character as a device and agent that filters the symbolic value of Argentinian and Latin American literature and influences the construction of literary taste in other communities and publishers in Latin America and Spain. All this accumulated capital (social, symbolic and economic), in which Eterna Cadencia has invested heavily, contributes to the visibility of the publisher and to that of the authors it publishes; who later appear in the catalogues of other independent European publishers or of the large conglomerates. But, principally, it influences the distribution of power and its decolonial change of meaning since, for the first time in more than half a century, a company with Latin American capital is playing a dominant role on the transatlantic independent book market. We cannot ignore the fact that we are talking about Argentina, one of the more culturally-developed countries in Latin America, with a robust publishing industry and a highly prestigious literary tradition. Eterna Cadencia is a medium-sized publishing house that puts great effort into the transnational marketing of its products. Even so its editorial policy is risky, since it focuses not only on the production of resistant aesthetics and languages, but on promoting and circulating essays in a variety of social spaces, such as the Eterna Cadencia bookshop itself, the FILBA Foundation and fair. With all this effort to widen its visibility and capital (social and symbolic, in this order), Eterna Cadencia has generated a community (Laclau, 1996; Nancy, 1990) of producers (with other independent publishers with similar aesthetics), young writers and readers who distribute, in the Southern Cone and in Spain, writing practices bordering on the avant-garde that are later assimilated by the large German and Spanish conglomerates. And, as our study shows, other hegemonies are possible in the independent Spanish-language publishing industry of the 21st century.

To conclude, as future lines of research, the substantial presence of publishing houses on Twitter opens the door to studies that focus on the forms of communication used and the ways to contact authors, readers and those interested in publishing, not only in independent firms but in the large groups. New avenues open up to articulate new cross-sectional research questions to analyze the book industry in current-day culture using online data and digital networks.